\documentstyle[prl,aps]{revtex}
\twocolumn

\begin{document}

\title{The Physicist's Guide to the Orchestra}

\author{Jean-Marc Bonard}

\address{D\'{e}partement de Physique, \'{E}cole Polytechnique 
F\'{e}d\'{e}rale de Lausanne,\\CH-1015 Lausanne EPFL, Switzerland \\
Email: jean-marc.bonard@epfl.ch}

\maketitle

\begin{abstract}
An experimental study of strings, woodwinds (organ pipe, flute, 
clarinet, saxophone and recorder), and the voice was undertaken to 
illustrate the basic principles of sound production in music 
instruments.  The setup used is simple and consists of common 
laboratory equipment.  Although the canonical examples (standing wave 
on a string, in an open and closed pipe) are easily reproduced, they 
fail to explain the majority of the measurements.  The reasons for 
these deviations are outlined and discussed.
\end{abstract}


\section*{Introduction}

Being a clarinet enthusiast, I wanted to share the richness of music 
(and the physicist's way of approaching it) to engineering students 
within the frame of their general physics classes.  The basics are 
reported in nearly every physics textbook and many details are 
provided in more specialized contributions 
\cite{leipp,benade,whites,rossing,fletross,taylor}.  I could however 
not find a comparison of the waveforms and spectra of different 
instruments in any of these sources.  I therefore measured and 
compared this data by myself.

The whole orchestra being too ambitious for a start, I restricted 
myself to two instrument families, namely strings and woodwinds (and 
the former are given a rather perfunctory treatment) and included the 
voice.  The waveform and spectra presented in the first part 
illustrate the behavior of standing waves on strings and in closed and 
open pipes (variations 1--2), as well as highlight the basic 
differences between the timbre of the instruments (variations 3--5).  
In the second part, we will note that although the instruments are 
easy to identify, variations between models or performers remain hard 
to assess (variation 6).  Furthermore, the simple models fail to 
explain the characteristics of some instruments (variations 6--8).

\section*{Theme}

Physicists have been studying the spectrum of musical instruments at 
least since the 1940s.  The first available methods were based on 
heterodyne analysis \cite{benadelarson} or on sonographs 
\cite{leipp,laloe}.  The development of the Fast Fourier Transform 
(FFT) algorithm coupled with the apparition of fast and relatively 
cheap microprocessors has greatly facilitated the task of the 
musically inclined physicists.  Quite sophisticated analysers have 
been realized \cite{brown} but setups based on commercial instruments 
work just as well for basic analysis \cite{miers,smedley}.

The waveforms have been acquired with the setup presented in 
Figure~\ref{setup}.  It consists of a condenser microphone connected 
directly to a 125~MHz LeCroy9400 digital oscilloscope.  The spectrum 
was calculated by the data acquisition program LabView by FFT (this 
task can be directly performed on most modern oscilloscopes).  On most 
figures, the time axis of the waveforms has been scaled and aligned 
for easier comparison.  The spectra are given over 10 or 20 harmonics 
with ticks corresponding to multiples of the fundamental frequency.

The waveforms and corresponding spectra analyzed here represent the 
continuous part of the sound only, which is only a small part of the 
character of a note.  The dependence on the type of attack, duration 
or strength was not considered, and would be a fascinating study in 
itself.  It is also clear that a more profound analysis would require 
a careful consideration of instrument and microphone position, as well 
as a calibration of the room acoustics and of the microphone response 
\cite{benadelarson}.  None of these points has been taken into 
account.

\begin{figure}
\caption{The experimental setup.}
\label{setup}
\end{figure}

\section*{Variation 1: vibrating strings}

A string instrument in its simplest form is composed of a stretched 
string over a resonance body that transmits the vibration of the 
string to the air.  Despite the simplicity of the vibrating system 
string instruments show a phenomenal diversity of timbre 
\cite{timbre}.  This arises from the variety of excitation as the 
string can be plucked (guitar, harp, harpsichord), bowed (violin, 
viola\ldots), or struck (piano).  The resonance body plays also a 
great role as is attested by the timbre difference between a guitar 
and a banjo.

When a continuous transverse wave is generated on a stretched string 
of length $l$, a standing wave forms following the superposition of 
the waves after reflection at the stops.  Simple considerations show 
that the only allowed modes of vibration correspond to wavelengths of 
$\lambda = 2l/n$ where $n \geq 1$ 
\cite{benade,whites,rossing,fletross}, which forms a harmonic series 
\cite{noteharmonic} (inset of Figure~\ref{pluckbow}).  The vibration 
of a string will be a superposition of the different modes with 
varying amplitudes determined by the mode of excitation, time since 
the excitation etc.

One of the most simple string instrument, the string sonometer, is 
formed of a simple stretched string over a box-shaped resonance body.  
Figure~\ref{pluckbow} shows the sound produced by a plucked and bowed 
sonometer tuned to A$_{2}$ \cite{notescales} and demonstrates the 
richness of the sound produced by a vibrating string as many intense 
upper harmonics are detected.  During bowing for example, the 5th, 
16th and 23rd harmonics are stronger than the fundamental.

The point and manner of excitation along the string (what physicists 
call the initial conditions) influences decisively the timbre: the 
spectra displayed in Figure~\ref{pluckbow} differ markedly, especially 
in the higher modes.  By plucking a string, we impose an initial 
condition such that the shape of the string is triangular with zero 
and maximal displacement at the stops and at the position of the 
finger, respectively.  The relation between position and intensity of 
the excited harmonics can be easily predicted (but is not easy to 
reproduce experimentally).  This relation is not as simple for the 
bowed string, since the bow imparts both displacement (a fraction of 
mm) and velocity ($\sim$0.5~m/s) \cite{broomfield}.

Finally, we always consider that the properties of the vibrating 
string are ideal.  A real string has however some stiffness, which 
causes an increase of the frequency of the higher modes with respect 
to an ideal string \cite{rossing}.  This can be detected on 
Figure~\ref{pluckbow}, especially for the plucked string (presumably 
because of the larger displacement).  As a consequence, the harmonics 
of a string are systematically sharp, be it for plucked 
\cite{smedley}, bowed \cite{halter} or struck strings such as in a 
piano \cite{brown,fletcher}.

\begin{figure}
\caption{Waveforms and corresponding spectra of a sonometer (A$_{2}$, 
110~Hz) plucked and bowed at $1/10$ of the string length, with in 
inset the first three vibration modes.  The thick line above the 
waveforms indicates the period of an oscillation.}
\label{pluckbow}
\end{figure}

\section*{Variation 2: vibrating air columns}

The principle of wind instruments is a bit more complicated than that 
of strings.  The vibrating medium is the air inside a pipe that acts 
as a resonator where a standing wave can form.  The ends of the pipe 
determine the boundary conditions.  At a closed end, the amplitude of 
vibration is zero and the pressure variation is maximal (the 
displacement of the air and the resulting variation of pressure are in 
anti-phase).  Conversely, the pressure will remain constant at the end 
of an open pipe and the standing wave shows a pressure node and a 
displacement antinode.  This is schematically shown in the insets of 
Figure~\ref{organ}.

As a consequence, a complete series of harmonics can form in a pipe of 
length $l$ open at both ends with wavelengths equal to $2 \cdot l / n$ 
with $n \geq 1$.  If the pipe is closed at one end, the wavelength of 
the fundamental corresponds to four times the length of the pipe and 
only the odd harmonics of wavelengths equal to $4 \cdot l/(2n+1)$ with 
$n \geq 0$ are allowed \cite{benade,whites,rossing,fletross}.

The vibration of the air can be excited by different means.  The most 
simple one is to produce an edge tone by steering an airjet over an 
edge [like the top of a bottle or the edge of the organ flue pipe 
shown in Figure~\ref{excite}(a)].  The edge forms an obstacle for the 
jet and generates periodic vortices at the mouth of the instrument.  
The vortices produce in turn periodic displacements of the air 
molecules.  When the edge forms the upper portion of a pipe, the edge 
tone is locked by resonance to the modes of the pipe.  The pitch can 
then only be changed by increasing the frequency of the vortices 
(i.e., by blowing faster) to lock the edge tone in a higher mode 
(which is exactly how flautists change the register and reach higher 
octaves with their instruments).  Such an edge excitation acts like an 
open end, as the vortices induce air displacement but no pressure 
variations.

The other means of excitation in wind instruments involve a mechanical 
vibration: that of the performer's lips for brass instruments or of a 
reed for woodwinds.  Similarly to the edge tones, the vibration of the 
lips or of the reed is locked to the resonances of the pipe.  The 
simple reed of the clarinet [see Figure~\ref{excite}(b)] and of the 
saxophone, and the double reed of the oboe and bassoon, acts actually 
as a pressure regulator by admitting periodically air packets into the 
pipe.  A reed is therefore equivalent to a closed end as it produces a 
pressure antinode.

\begin{figure}
\caption{Excitation systems for woodwinds: (a) the edge of a flue 
organ pipe, and (b) the reed and mouthpiece of a clarinet.}
\label{excite}
\end{figure}

\begin{figure}
\caption{Waveform and corresponding spectra of a closed and open organ 
flue pipe (B$\flat_{3}$, 235~Hz and B$\flat_{4}$, 470~Hz, 
respectively).  The timescale of the upper waveform has been divided 
by two with respect to the lower waveform.  The insets show the first 
three vibration modes for the variation of the pressure.}
\label{organ}
\end{figure}

We can verify the above principles with a square wooden flue organ 
pipe of 0.35~m length.  The excitation system is reproduced in 
Figure~\ref{excite}(a) and acts as an open end, and the other end can 
be either closed or open.  As shown on Figure~\ref{organ}, the 
fundamental of the closed pipe is found at 235~Hz, which corresponds 
well to a wavelength of $\lambda = v/f = 4 \cdot 0.35 = 1.4$~m with $v 
= 330$~m/s.  The waveform is nearly triangular, and the even harmonics 
are far weaker than the odd.  The same pipe with its end open sounds 
one full octave higher (the wavelength of the fundamental is shorter 
by a factor of 2) and displays a complete series of harmonics.

\section*{Variation 3: tutti}

\begin{figure}
\caption{Waveform of a violin, recorder, flute, clarinet, saxophone 
and of the author singing the french vowel ``aa'' (as in sat) 
(A$_{4}$, 440~Hz for the former and A$_{3}$, 220~Hz for the latter 
three instruments).  The timescale of the upper waveforms has been 
divided by two with respect to the lower waveforms.}
\label{compwf}
\end{figure}

\begin{figure}
\caption{Spectra corresponding to the waveforms of 
Figure~\ref{compwf}.}
\label{compfft}
\end{figure}

We are now ready to study the behaviour of most strings and woodwinds.  
Figures~\ref{compwf} and \ref{compfft} show the waveforms and spectra 
of six different instruments.  Table~\ref{t1} also summarizes the 
characteristics of the woodwinds studied here.  A quick glance shows 
numerous disparities between the instruments, and we will try now to 
understand these timbre variations and their origin in more detail.

\begin{table}
\caption{Characteristics of the woodwinds studied in this work.}
\vspace{0.2 cm}
\begin{tabular}{@{}l l l}
instrument & bore & excitation\\
\hline
flute & cylindrical & edge \\
clarinet & cylindrical & single reed \\
saxophone & conical & single reed \\
recorder & cylindrical & edge \\
\label{t1}
\end{tabular}
\end{table}

\section*{Variation 4: the violin}

The violin produces a very rich sound with at least 20 strong 
harmonics and complex waveforms, as was the case for the string 
sonometer in Figure~\ref{pluckbow}.  The strongest mode is not the 
fundamental, but the 7th harmonic in the case of Figure~\ref{compfft}.

\section*{Variation 5: the flute}

As can be seen on Figure~\ref{compwf}, woodwinds show simple waveforms 
and spectra when compared to string instruments.  The flute (flauto 
traverso) is a textbook example of an wind instrument with open ends 
as the pipe is (nearly) cylindric over the whole length.  The most 
salient feature of the flute is the limited number of harmonics 
($\sim$7) with an intensity that decreases monotonously 
\cite{klein,smith}.  The timbre is also very similar for the first two 
registers (not shown here).

\section*{Variation 6 (menuetto): the clarinet}

We have seen that a pipe closed at one end shows only odd harmonics: 
the clarinet, with its simple reed and (nearly) cylindric bore, should 
be a prototype of such a pipe.

At first sight, this is indeed the case.  In the low register 
(Figure~\ref{compwf} for an A$_{3}$ \cite{note2}), the odd harmonics 
are clearly the strongest modes.  The even harmonics, although 
present, are strongly attenuated (at least up to the 6th harmonic).  
There are other marked differences with the flute.  First, the sound 
is far richer in higher harmonics.  Second, the waveform varies 
considerably with the pitch as displayed on Figure~\ref{clarinetLa}.  
The contents of higher harmonics strongly decreases from 20 for the 
A$_{3}$, to 9 and 5 for the A$_{4}$ and A$_{5}$.  The contribution of 
the even harmonics becomes also increasingly important.  The third 
mode remains more intense than the second for the A$_{4}$, but this is 
not the case anymore for A$_{5}$.

\begin{figure}
\caption{Waveform and corresponding spectra of a clarinet (A$_{3}$, 
A$_{4}$, A$_{5}$ at 220, 440 and 880~Hz, respectively).  The timescale 
of the second and third waveform have been divided by two and four 
with respect to the lower waveform.}
\label{clarinetLa}
\end{figure}

The clarinet shows thus a fascinating behavior: it responds like a 
pipe closed at one end in the lower register but gives a sound with 
strong even harmonics in the higher registers.  The timbre varies 
therefore as the pitch is increased, with a very distinctive sound for 
each register.  This is due to several facts.  First, the bore of the 
clarinet is not perfectly cylindric but has tapered and slightly 
conical sections \cite{benade}.  Second, the flared bell, the 
constricting mouthpiece [Figure~\ref{excite}(b)] and the toneholes 
(even if they are closed) perturb significantly the standing waves.  
Finally, for wavelengths comparable to the diameter of the toneholes, 
the sound wave is no longer reflected at the open tonehole but 
continues to propagate down the pipe.  This corresponds a frequency of 
$\sim$1500~Hz in typical clarinets \cite{benade,benadekou}, and the 
sound will show increasing amounts of even harmonics with increasing 
pitch, as found on Figure~\ref{clarinetLa}.

Figure~\ref{clarinetLa} leaves out one important feature.  The 
clarinet does not change from the first to the second register by an 
octave (i.e., by doubling the frequency), but by a duodecime (tripling 
the frequency).  This feature is due to the excitation system alone 
(the reed acts as a closed end), as can be easily demonstrated by 
replacing the mouthpiece of a flute (or of a recorder) with a clarinet 
mouthpiece mounted on a section of pipe such that the overall length 
of the instrument remains identical.  The instrument sounds a full 
octave lower and changes registers in duodecimes, not in octaves.  The 
reverse effect can be demonstrated by mounting a flute mouthpiece on a 
clarinet.

\subsection*{Trio I: timbre quality}

It appears from Figure~\ref{compwf} that it is quite easy to recognize 
an instrument family by its waveform or spectra.  It would be 
tantalizing if one could also recognize one clarinet from another, for 
example to choose and buy a good instrument.

Figure~\ref{diffclar} shows the spectra of my three clarinets playing 
the same written note \cite{note2}, with the same mouthpiece, reed, 
embouchure and loudness.  The upper curve correspond to my first 
instrument, a cheap wooden B$\flat$ student model.  The two lower 
curves were obtained with professional grade B$\flat$ and A clarinets.  
I can identify each instrument by playing a few notes from the 
produced sound and from muscular sensations in the embouchure and 
respiratory apparatus.

At first glance, the spectra of the three clarinets are readily 
comparable.  Closer inspection shows that the spectra begin to differ 
from the 10th harmonic on!  There are actually far less variations in 
relative intensities between the two B$\flat$ instruments than between 
the two pro clarinets.  The pro B$\flat$ seems to be slightly richer 
in harmonics than the student model.  The A has no strong harmonics 
beyond the 11th.  This leads to the conclusion that the B$\flat$ and A 
clarinets are (slightly) different instruments (many clarinetists will 
agree with that point).  The measured differences between two B$\flat$ 
clarinets remain however quite subtle despite the huge and easily 
audible difference in timbre.

\begin{figure}
\caption{Spectra of the written C$_{4}$ of a student B$\flat$ and a 
professional grade B$\flat$ and A clarinet played with the same 
mouthpiece and reed (sounding B$\flat_{3}$, 235~Hz, and A$_{3}$, 
220~Hz, respectively).}
\label{diffclar}
\end{figure}

\subsection*{Trio II: tone quality}

Is it possible to tell apart a good from a bad tone?  This question is 
of utmost importance for every musician to obtain the desired tone 
quality.  Figure~\ref{clarinettone} shows two clarinet tones obtained 
on the same instrument, with the same mouthpiece and reed.  The first 
is a good tone: one could describe it as fullbodied, agreeable to the 
ear.  The second is a beginner's tone: emitted with a closed throat 
and weak.  The difference is instantly audible but difficult to 
quantify from Figure~\ref{clarinettone}.  The variations appear again 
in the higher harmonics: the bad tone show no harmonics beyond the 
12th, which is at least five modes less than the good tone.

\begin{figure}
\caption{Spectra of a good and of a bad sound on the clarinet 
(B$\flat_{3}$, 235~Hz).}
\label{clarinettone}
\end{figure}

It is quite astonishing that the quality of the sound is determined by 
the presence (or absence) of high harmonics with amplitudes that are 
at least 40~dB (a factor 10$^4$) weaker than the fundamental!  
Musicians are sensitive to very subtle effects which are difficult to 
(a) link to a physically measurable value and (b) to quantify 
precisely.  Conventional statistics have proven ineffective for 
classifying the sound quality: interestingly, effective solutions 
based on neural networks have been recently demonstrated \cite{fasel}.

\section*{Variation 7: the saxophone}

Can one predict the spectra of the saxophone, a single reed instrument 
with a truncated conical bore, by extrapolation from the previous 
observations?  The sax is a wind instrument, which would imply a 
limited number of harmonics, and a spectra mainly composed of odd 
harmonics because of the reed.  A short glance at Figures~\ref{compwf} 
and \ref{compfft} shows that both predictions are wrong.  The even 
harmonics are as strong as the odd \cite{benadesax}.  The sound 
remains very rich in harmonics even in the higher registers, far more 
than for the clarinet, and the timbre changes only slightly between 
the first and the second register.  The saxophone does not behave at 
all like a clarinet!

The main reason is the form of the bore: in a cone, the standing waves 
are not plane but spherical \cite{benade,benadesax,ayers,boutillon}.  
This has profound implications for the standing wave pattern 
\cite{ayers}.  In short, the intensity of a wave travelling down or up 
the pipe is in first approximation constant along the pipe, which 
implies that the amplitude scales with the inverse of the distance to 
the cone apex.  The waves interfer to form a spherical standing wave 
with pressure nodes separated by the usual half-wavelength spacing, 
but with an amplitude that varies as the inverse of the distance to 
the cone apex.  This is true for a closed as well as an open end 
\cite{ayers}.  A conical bore shows therefore a complete harmonic 
series, be it excited with a reed, the lips or an edge 
\cite{boutillon}!  It would seem also that the conical pipe of the 
saxophone favors the higher harmonics as compared to the cylindric 
bore of the clarinet.

\section*{Variation 8: the recorder}

The predictions for the saxophone were wrong, so let's try again with 
another instrument -- the recorder for example.  That should be easy: 
the bore is nearly cylindrical, it is excited by an edge and should 
therefore be similar to the open organ flue pipe.  I expected a 
limited number of harmonics and a full harmonic series.  
Figure~\ref{compwf} shows that I was wrong again and this puzzled me 
greatly.  The alto recorder indeed has a limited number of harmonics, 
and a similar timbre in the two registers.  But it shows the spectrum 
of a {\it closed} pipe -- the even harmonics are more suppressed than 
for the clarinet -- and despite that it changes registers in {\it 
octaves}!

What is the explanation for the odd behaviour of the recorder?  The 
player generates an airjet by blowing into a rectangular windcanal, 
which is then cut by the edge (see Figure~\ref{excite}).  It appears 
from calculations that the position of the edge relative to the airjet 
influences critically the intensity of the different harmonics 
\cite{fletdou}.  When the edge cuts the side of the jet, the full 
harmonic series is observed.  The even harmonics are however 
completely absent when the edge is positioned in the center of the 
jet, as is the case for most modern recorders (among those the one I 
used).  This of course does not affect the modes of resonance of the 
instrument: the second harmonic can be excited easily by increasing 
the speed of the airjet, which raises the pitch by an octave.  It 
follows also that I have been very lucky with the open organ flue pipe 
-- which follows the expected behaviour shown on Figure~\ref{organ} 
thanks to a favorable position of the edge with respect to the airjet 
\cite{fletdou}!

\section*{Variation 9: a capella}

We perform frequently with a peculiar and versatile musical 
instrument, namely our voice.  Few instruments have such varied 
expressive possibilities and ability to change the timbre and 
loudness.  From the point of view of musical acoustics, the voice is a 
combination of a string and a wind instrument.  A pipe, the vocal 
tract, is excited by the vibration of the vocal cords that generate a 
complete harmonic series as is usual for vibrating strings (see 
Figure~\ref{compfft}).  The timbre is however determined by the shape 
of the vocal tract that acts as a resonator.  Depending on the 
position of the tongue and on the mouth opening, the position and 
width of the formants of the vocal tract (the broad resonances, 
indicated in Figure~\ref{voice}) can be varied and some harmonics 
produced by the vocal cords are favored with respect to others 
\cite{whites,sundberg}.  Note that vocal tract and vocal cords are 
independent of each other, which implies that the timbre of the voice 
will change with the pitch for a given tract shape as the harmonics 
are shifted towards higher frequencies while the position of the 
formants remains constant.

The effect of the vocal tract shape is displayed in Figure~\ref{voice} 
for three vowels sung at the same pitch (the formants are also 
indicated).  The tongue is placed closed to the palate to produce the 
``ii'': it is nearly sinusoidal with weak upper harmonics.  The first 
formant peaks around 200~Hz and decreases rapidly.  The second and 
third formant around 2000 and 3000~Hz are however easily visible.  The 
``ou'' results from a single formant with a maximum around 300~Hz and 
a slowly decreasing tail: the waveform is more complex and richer in 
higher harmonics, giving a flute-like sound.  The ``aa'' is obtained 
with an open tract and is far more complex.  The most intense harmonic 
is the third because of the relatively high position ($\sim$800~Hz) 
and large width of the first formant.  The second and the third 
formant are as intense as the first and give a significant amount of 
higher harmonics to the sound.

\begin{figure}
\caption{Waveform and corresponding spectra of the author singing an 
A$_{3}$ (220~Hz) on three different vowels: the french ``ii'' (as in 
this), ``ou'' (as in shoe) and ``aa'' (as in sat).  The formants are 
indicated for each spectrum by a dotted line in linear scale.}
\label{voice}
\end{figure}

\section*{Finale}

We have seen that the physicist's approach to musical instruments 
opens fascinating and complex possibilities.  The classical examples 
(closed and open pipe, for example) are easy to reproduce, but one 
steps quickly into territory uncharted by the classical physics 
textbook, which makes the exploration all the more exciting.  It 
remains also that instruments are easy to identify by their timbre, 
but that it is quite difficult to tell two different models from one 
another and to classify the quality of the produced sound.  It may be 
even more difficult (not to say impossible) to examine the quality of 
an interpretation and to understand why well-played music touches us 
so deeply.

Musical acoustics is a beautiful subject to teach at every level.  
Music appeals to everybody and a lot of students play or have played 
at some stage an instrument: this makes often for lively 
demonstrations in front of the class.  It involves both wave mechanics 
and fluid mechanics in quite complex ways, and a simple experimental 
setup can offer direct and compelling insights in the physics of sound 
production.  I hope that this excursion in the basic physics of 
musical instruments will motivate some of the readers to include the 
subject in their curriculum and that it may provide helpful material 
for those who already do.

\section*{Acknowledgments}

I thank heartily the different people that either lent me their 
instrument or that took some time to come and play in the lab: Ariane 
Michellod (flute), S{\'e}verine Michellod (recorders), Stephan Fedrigo 
(violin -- handcrafted by the performer!)  and Lukas B{\"u}rgi 
(saxophone).  I am also greatly indebted to Paul Braissant, Bernard 
Egger and Yvette Fazan, who maintain and expand an impressive 
collection of physics demonstration experiments at EPFL, and who are 
never put off by the sometimes strange requests of physics teachers.


\begin{references}
\bibitem{timbre}
The timbre is usually defined as the quality of tone distinctive of a 
particular singing voice or musical instrument, and is essentially 
determined by the relative intensity of the different harmonics 
\cite{noteharmonic}.
\bibitem{noteharmonic}
The frequency ratios given by the modes of vibration on a string are 
pleasing to the ear and form the basis of western music (e.g., between 
the first and second mode, 2:1 or an octave; the second and third 
mode, 3:2 or a fifth).  The series of modes with wavelengths $\lambda 
= 2l/n$ and corresponding freqencies $f = v/(2l/n)$ (where $v$ is the 
velocity of the wave) is hence called the harmonic series, and the 
term mode and harmonic is used interchangeably.
\bibitem{notescales}
The names of the notes refer to the tempered scale, formed of 12 
semitones per octave with a frequency ratio between each semitone of 
$\sqrt[12]{2} = 1.059464$.  The A$_{4}$ is defined as the tuning A at 
440~Hz, and the index of the note gives the octave.  For further 
information on scales and intervalls see, e.g., Ref.~\cite{whites}.
\end{references}
\end{document}